\documentclass[letterpaper]{jpconf}
\usepackage{graphicx}

\begin{document}
\title{Baryon chiral perturbation theory transferred to hole-doped
antiferromagnets on the honeycomb lattice}

\author{F-J Jiang$^1$, F K\"ampfer$^2$, B Bessire$^3$, M Wirz$^4$, C P
Hofmann$^5$, and U-J Wiese$^6$ }

\address{$^1$ Department of Physics, National Taiwan Normal University, 88,
Sec.\ 4, Ting-Chou Rd.\, Taipei 116, Taiwan}
\address{$^2$ BKW FMB Energy Ltd, Energy Trading Unit, 3000 Bern, Switzerland}
\address{$^3$ Institute of Applied Physics, Bern University, Sidlerstrasse 5,
CH-3012 Bern, Switzerland}
\address{$^4$ Mathematical Institute, Bern University, Sidlerstrasse 5,
CH-3012 Bern, Switzerland}
\address{$^5$ Facultad de Ciencias, Universidad de Colima, Bernal D\'iaz del
Castillo 340, Colima C.P.\ 28045, Mexico}
\address{$^6$ Institute for Theoretical Physics, Albert Einstein Center for
Fundamental Physics, Bern University, Sidlerstrasse 5, CH-3012 Bern,
Switzerland}

\ead{christoph.peter.hofmann@gmail.com}

\begin{abstract}
A systematic low-energy effective field theory for hole-doped antiferromagnets
on the honeycomb lattice is constructed. The formalism is then used to
investigate spiral phases in the staggered magnetization as well as the
formation of two-hole bound states.

\end{abstract}

\section{Motivation}

Although the phenomenon of high-temperature superconductivity was discovered
more than twenty-five years ago \cite{Bed86}, the dynamical mechanism behind
it still remains a mystery today. The reason may be readily identified: Due to
the nonperturbative nature of the problem, analytic studies usually suffer
from uncontrolled approximations, while numerical simulations of the
microscopic Hubbard and $t$-$J$-type models suffer from a severe fermion sign
problem away from half-filling.

In the present work, we will address the physics of the antiferromagnetic
precursors of high-temperature
superconductivity from a universal and model-independent point of view, based
on the method of effective Lagrangians. The effective Lagrangian method is
very well-established in particle physics, where chiral perturbation theory
(CHPT) \cite{Wei79,Gas85} represents the effective theory of quantum
chromodynamics (QCD). It operates in the Goldstone boson sector of QCD, taking
into account the octet of the pseudoscalar mesons that emerge due to the
spontaneously broken chiral symmetry. Baryon chiral perturbation theory
\cite{Gas88,Jen91,Ber92,Bec99} is an extension of CHPT as it also includes
heavy particles such as the baryon octet or decuplet, i.e.\ those degrees of
freedom that remain massive in the chiral limit.

Spontaneous symmetry breaking is also a very common phenomenon in condensed
matter physics: Magnons e.g.\ are the Goldstone bosons of a spontaneously
broken spin symmetry $SU(2)_s \to U(1)_s$, while phonons emerge due to a
spontaneously broken translation symmetry. For these systems effective
Lagrangians have been constructed in Refs.~\cite{Has93,Leu94,Leu97,RS99a}
and various applications have been considered in
Refs.~\cite{Hof99a,Hof99b,RS99b,RS00,Hof02,Hof10,Hof11}. 

However, the incorporation of heavy degrees of freedom in the condensed matter
domain -- in a transparent and systematic manner -- has only been achieved
recently. An effective theory for weakly doped antiferromagnets, the analog of
baryon chiral perturbation theory, has been constructed for hole- and
electron-doped antiferromagnets on the square lattice in
Refs.\cite{Kae05,Bru06,Bru07b} and for hole-doped antiferromagnets on the
honeycomb lattice in Ref.\cite{KBWHJW12}. Further applications of the
formalism were discussed in Refs.\cite{Bru06a,Bru07a,JKHW09}, demonstrating
that the effective Lagrangian method allows one to gain insight into the
physics of these insulating precursors of high-temperature superconductors in
a systematic and unambiguous manner.

In this presentation we will focus on hole-doped antiferromagnets on a
honeycomb lattice. We first review the systematic construction of the
effective Lagrangian for magnons and holes for these systems and will then
consider two applications of the formalism: The emergence of spiral phases in
the staggered magnetization order parameter as well as the formation of
two-hole bound states mediated by magnon exchange.

We would also like to point out that in a recent article on an analytically
solvable microscopic model for a hole-doped ferromagnet in 1+1 dimensions
\cite{GHKW10}, the correctness of the effective field theory approach was
demonstrated by comparing the effective theory predictions with the
microscopic calculation. Likewise, in a series of high-accuracy investigations
of the antiferromagnetic spin-$\frac{1}{2}$ quantum Heisenberg model on a
square lattice using the loop-cluster algorithm
\cite{WJ94,GHJNW09,JW11,GHJPSW11},
the Monte Carlo data were confronted with the analytic predictions of magnon
chiral perturbation theory and the low-energy constants were extracted with
permille accuracy. All these tests unambiguously demonstrate that the
effective Lagrangian approach provides a rigorous and systematic derivative
expansion for both ferromagnetic and antiferromagnetic systems.

\section{Construction of the effective field theory for holes and magnons}

The effective Lagrangian approach is based on a symmetry analysis of the
underlying theory, i.e., in the present case the Hubbard or $t$-$J$-type
models which are believed to be minimal models for high-temperature
superconductors \cite{And87}. We thus first perform a rigorous symmetry
analysis of these models in order to then construct the effective Lagrangian
for holes and magnons.

\subsection{Symmetry analysis}

Let $c_{x s} ^\dagger$ denote the operator which creates a fermion with spin
$s\in\lbrace\uparrow,\downarrow\rbrace$ on a lattice site
\mbox{$x=(x_1,x_2)$}. The corresponding annihilation operator is $c_{xs}$.
These fermion operators obey the canonical anticommutation relations
\begin{equation}
\label{commutators1}
\{c_{xs}^{\dagger},c_{ys'}\} = \delta_{xy} \delta_{ss'}, \ 
\{c_{x s},c_{y s'}\} = \{c_{x s}^{\dagger},c_{y s'}^{\dagger} \} = 0.
\end{equation}
The second quantized Hubbard Hamiltonian is defined by
\begin{equation}\label{hubh}
H = - t \sum_{{\langle x, y\rangle};\,{s=\uparrow ,\downarrow}}
(c_{xs} ^\dagger c_{ys} + c_{ys} ^\dagger c_{xs})
+ U\sum_x c_{x\uparrow}^\dagger c_{x\uparrow}c_{x\downarrow}^\dagger c_{x\downarrow}
- \mu ' \sum_{{x};\,{s=\uparrow ,\downarrow}} c_{xs}^\dagger c_{xs},
\end{equation}
where $\langle x, y\rangle$ indicates summation over nearest neighbors, $t$ is
the hopping parameter, and the parameter $U>0$ fixes the strength of the
Coulomb repulsion between two fermions located on the same lattice site. The
parameter $\mu'$ denotes the chemical potential.

Using the fermion creation and annihilation operators, we define the following
$SU(2)_s$ Pauli spinors
\begin{equation}\label{pspinor}
c_x^\dagger = \left(c_{x \uparrow}^\dagger, c_{x \downarrow}^\dagger\right), \qquad
c_x = \left(\begin{array}{c} c_{x \uparrow} \\ c_{x \downarrow} 
\end{array} \right).
\end{equation}
In terms of these operators, the Hubbard model can be reformulated as
\begin{equation}\label{hubhspin}
H = - t \sum_{\langle x y\rangle} (c_x^\dagger c_y + c_y^\dagger c_x) + 
\frac{U}{2} \sum_x (c_x^\dagger c_x - 1)^2 - \mu \sum_x (c_x^\dagger c_x - 1).
\end{equation}
The parameter $\mu = \mu' - \frac{U}{2}$ controls doping where the fermions
are counted with respect to half-filling. 

Since all terms in the effective Lagrangian must be invariant under all
symmetries of the Hubbard model, a careful symmetry analysis of
Eq.(\ref{hubhspin}) is needed. Let us divide the symmetries of the Hubbard
model into two categories: Continuous symmetries, which are internal
symmetries of Eq.(\ref{hubhspin}), and discrete symmetries, which are symmetry
transformations of the underlying honeycomb lattice depicted in figure 1.

\begin{figure}
\begin{center}
\includegraphics[scale=0.7]{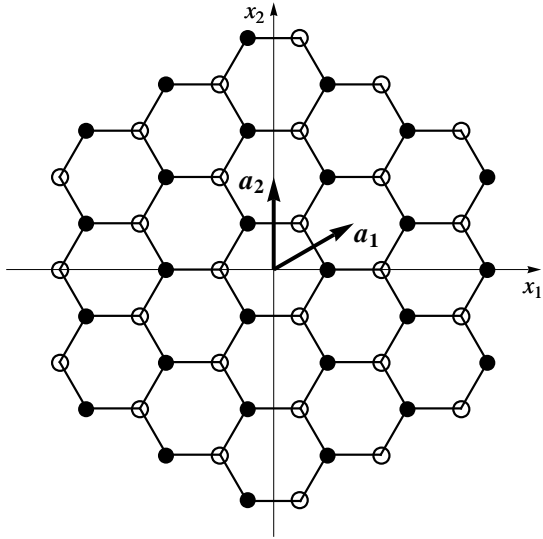}
\end{center}
\caption{\it Bipartite non-Bravais honeycomb lattice consisting of two
triangular Bravais sublattices. The translation vectors are $a_1$ and $a_2$.}
\end{figure}

The continuous symmetries are the $SU(2)_s$ spin rotation symmetry, the
$U(1)_Q$ fermion number symmetry and its non-Abelian extension $SU(2)_Q$. The
discrete symmetries include translations ($D_i$) along the two primitive
lattice vectors of the honeycomb lattice, rotations ($O$) by 60 degrees around
the center of a hexagon, and reflections ($R$) at the $x_1$-axis going through
the center of the hexagon. There is also time reversal which is implemented by
an anti-unitary operator $T$. This symmetry will be considered later on in the
effective field theory framework. Furthermore, in the construction of the
effective field theory for magnons and holes, it will turn out to be useful to
incorporate the combined symmetry $O'$ consisting of a spatial rotation $O$
and a global $SU(2)_s$ spin rotation $g = i\sigma_2$. Likewise, we define the
composed transformation $T'$, consisting of a regular time-reversal $T$ and
the specific spin rotation $g = i \sigma_2$.

In \cite{Zha90,Yan90}, Yang and Zhang proved the existence of a non-Abelian
extension of the $U(1)_Q$ fermion number symmetry in the half-filled Hubbard
model. This pseudospin symmetry contains $U(1)_Q$ as a subgroup. The $SU(2)_Q$
symmetry is realized on the square as well as on the honeycomb lattice and is
generated by the three operators
\begin{eqnarray}
Q^+ & = & \sum_x (-1)^x c_{x \uparrow}^\dagger c_{x \downarrow}^\dagger, \qquad
Q^- = \sum_x (-1)^x c_{x \downarrow} c_{x \uparrow}, \nonumber \\
Q^3 & = &\sum_x \frac{1}{2}(c_{x \uparrow}^\dagger c_{x \uparrow} +
c_{x \downarrow}^\dagger c_{x \downarrow} - 1) = \frac{1}{2} Q.
\end{eqnarray}
The factor $(-1)^x $ distinguishes between the two triangular sublattices $A$
and $B$ of the honeycomb lattice. Defining $Q^1$ and $Q^2$ through
$Q^\pm = Q^1 \pm i Q^2$, one readily shows that the $SU(2)_Q$ Lie-algebra
$[Q^a,Q^b] = i \varepsilon_{abc} Q^c$, with $a, b, c \in \lbrace1,2,3\rbrace$,
indeed is satisfied  and that $[H,\vec Q]=0$ with
$\vec Q = (Q^1,Q^2,Q^3)$ for the Hubbard Hamiltonian with $\mu = 0$.

In order to write the Hubbard Hamiltonian Eq.(\ref{hubh}) or
Eq.(\ref{hubhspin}) in a manifestly invariant form under
\mbox{$SU(2)_s \times SU(2)_Q$}, we arrange the fermion operators in a
$2\times 2$ matrix-valued operator, arriving at the fermion representation
\begin{equation}\label{Coperator}
C_x = \left(\begin{array}{cc} c_{x \uparrow} &
(-1)^x \ c^\dagger_{x \downarrow} \\ c_{x \downarrow} &
- (-1)^x c_{x \uparrow}^\dagger \end{array} \right).
\end{equation}
This allows us to write down the Hubbard Hamiltonian in the manifestly
$SU(2)_s$, $U(1)_Q$, $D_i$, $O$, $O'$ and $R$ invariant form
\begin{equation}\label{HF}
H = - \frac{t}{2} \sum_{x, i} \mbox{Tr}[C_x^\dagger C_{x+\hat i} +
C_{x+\hat i}^\dagger C_x] + 
\frac{U}{12} \sum_x \mbox{Tr}[C_x^\dagger C_x C_x^\dagger C_x] -
\frac{\mu}{2} \sum_x \mbox{Tr}[C_x^\dagger C_x \sigma_3].
\end{equation}
The $\sigma_3$ Pauli matrix in the chemical potential term prevents the
Hubbard Hamiltonian from being invariant under $SU(2)_Q$ away from
half-filling. For $\mu \neq 0$, $SU(2)_Q$ is explicitly broken down to its
subgroup $U(1)_Q$. In addition, the pseudospin symmetry is realized in
Eq.(\ref{HF}) only for nearest-neighbor hopping. As soon as
next-to-nearest-neighbor hopping is included, the $SU(2)_Q$ invariance gets
lost even for $\mu=0$. The continuous $SU(2)_Q$ symmetry contains a discrete
particle-hole symmetry. Although this pseudospin symmetry is not present in
real materials, it will play an important role in the construction of the
effective field theory. The identification of the final effective fields for
holes will lead us to explicitly break the $SU(2)_Q$ symmetry in subsection
\ref{holes}.

\subsection{Effective field theory for magnons}

In this section we review the construction of the effective theory for
antiferromagnetic magnons on the honeycomb lattice. The basic object in the
effective theory is the staggered magnetization order parameter of the
antiferromagnet, which is described by a unit-vector field
\begin{equation}
\vec e(x) =
(\sin\theta(x) \cos\varphi(x),\sin\theta(x) \sin\varphi(x),\cos\theta(x)),
\end{equation}
in the coset space $SU(2)_s/U(1)_s = S^2$, where $x = (x_1,x_2,t)$ denotes a
point in $(2+1)$-dimensional space-time. A key ingredient for constructing the
effective field theory is the nonlinear realization of the global $SU(2)_s$
spin symmetry which is spontaneously broken down to its $U(1)_s$ subgroup,
which is discussed in detail in Ref.~\cite{Kae05}.

It turns out that the above parametrization of the magnon degrees of freedom
through the vector $\vec e(x)$ is not appropriate for this construction.
Rather, one uses composite magnon fields $v_\mu(x)$ whose components will be
used to couple the magnons to the fermions. The composite magnon field is
defined by
\begin{equation}
\label{defvmu}
v_\mu(x)= u(x) {\partial}_\mu u(x)^{\dagger},
\end{equation}
where the matrix $u(x)$ is related to the original magnetization vector
${\vec e}(x)=(e_1(x),e_2(x),e_3(x))$ by
\begin{eqnarray}
\label{umatrixdef}
u(x) & = \frac{1}{\sqrt{2(1+e_3(x))}}
\left(\begin{array}{cc} 1 + e_3(x) & e_1(x) - i e_2(x) \\
- e_1(x) - i e_2(x) & 1 + e_3(x) \end{array} \right) \nonumber \\[1ex]
& = \left(\begin{array}{cc}
\cos\left(\frac{\theta(x)}{2}\right) & \sin\left(\frac{\theta(x)}{2}\right)
\exp(-i\varphi(x))
\\[0.5ex]
-\sin\left(\frac{\theta(x)}{2}\right)\exp (i\varphi(x)) & 
\cos\left(\frac{\theta(x)}{2}\right)
 \end{array} \right).
\end{eqnarray}
The coupling of magnons to holes is then realized through the matrix-valued
anti-Hermitean field
\begin{equation}
v_\mu(x) = i v_\mu^a(x) \sigma_a, \qquad 
v_\mu^\pm(x) = v_\mu^1(x) \mp i v_\mu^2(x),
\end{equation}
which decomposes into an Abelian "gauge" field $v_\mu^3(x)$ and two vector
fields $v_\mu^\pm(x)$ "charged" under the unbroken subgroup $U(1)_s$. Here
$\vec \sigma$ are the Pauli matrices. These fields have a well-defined
transformation behavior under the symmetries which the effective theory
inherits from the underlying microscopic $t$-$J$ model:
\begin{eqnarray}
\label{symm_magnonfields}
SU(2)_s:& \quad & v_\mu(x)' = h(x) (v_\mu(x) + {\partial}_\mu) h(x)^\dagger,
\nonumber \\
D_i:& \quad & ^{D_i}v_\mu(x) = v_\mu(x),
\nonumber\\
O:& \quad & ^Ov_1(x) =
\tau(Ox)\Big\{ \mbox{$\frac{1}{2}$} v_1(Ox)+ \mbox{$\frac{\sqrt{3}}{2}$}
v_2(Ox) +  \mbox{$\frac{1}{2}$}{\partial}_1
+ \mbox{$\frac{\sqrt{3}}{2}$}{\partial}_2 \Big\} \tau(Ox)^\dagger,
\nonumber\\
& \quad & ^Ov_2(x) =
\tau(Ox)\Big\{ - \mbox{$\frac{\sqrt{3}}{2}$}v_1(Ox)
+ \mbox{$\frac{1}{2}$}v_2(Ox)
- \mbox{$\frac{\sqrt{3}}{2}$}{\partial}_1
+ \mbox{$\frac{1}{2}$}{\partial}_2 \Big\} \tau(Ox)^\dagger,
\nonumber\\
& \quad & ^Ov_t(x) =
\tau(Ox)(v_t(Ox)+{\partial}_t)\tau(Ox)^\dagger,\nonumber\\
R:&\quad & ^Rv_1(x) = v_1(Rx), \quad
^Rv_2(x) = -v_2(Rx), \nonumber \\
& \quad & ^Rv_t(x) = v_t(Rx),\nonumber \\
T:& \quad & ^Tv_i(x) = \tau(Tx)(v_i(Tx)+{\partial}_i)\tau(Tx)^\dagger,
\nonumber\\
& \quad & ^Tv_t(x) = -\tau(Tx)(v_t(Tx)+{\partial}_t)\tau(Tx)^\dagger.
\end{eqnarray}
In the above symmetry transformations, we have introduced the matrix
$\tau(x)$,
\begin{equation}
\tau(x)=
\left(
\begin{array}{cc}
0 & -\exp(-i \varphi(x))\\
\exp(i\varphi(x)) & 0
\end{array}
\right).
\end{equation}
Finally, the Abelian ``gauge'' transformation
\begin{equation}
h(x) = \exp(i\alpha(x)\sigma_3)
\end{equation}
belongs to the unbroken $U(1)_s$ subgroup of $SU(2)_s$ and acts on the
composite vector fields as
\begin{eqnarray}
v_\mu^3(x)' &=& v_\mu^3(x) - {\partial}_\mu \alpha(x), \nonumber \\ 
v_\mu^\pm(x)' &=& v_\mu^\pm(x) \exp(\pm 2 i \alpha(x)).
\end{eqnarray}
The magnon action, in terms of the composite magnon field $v_{\mu}(x)$, can
now be expressed as
\begin{equation}
\label{vmagnonaction}
S[v_\mu^\pm] = \int d^2\!x \, dt \ 2\rho_s \left(v^+_i v^-_i + \frac{1}{c^2}
v_t^+ v_t^-\right).
\end{equation}
Although the expression $v_{\mu}^{+}v_{\mu}^{-}$ looks like a mass term of a
charged vector field, it is just the kinetic term of a massless Goldstone
boson, since it contains derivatives acting on $u(x)$.

\subsection{Effective field theory for holes and magnons}
\label{holes}

Analytic calculations as well as Monte Carlo simulations in $t$-$J$-like
models on the honeycomb lattice have revealed that at small doping holes occur
in  pockets centered at lattice momenta $k^\alpha = - k^\beta
= (0,\frac{4\pi}{3\sqrt{3}a})$, and their copies in the periodic Brillouin
zone \cite{Lus04,Jia08}. The honeycomb lattice, illustrated in figure 1, is a
bipartite non-Bravais lattice which consists of two triangular Bravais
sublattices. The corresponding Brillouin zone and the location of the
corresponding hole pockets are shown in figure 2.
\begin{figure}
\begin{center}
\includegraphics[scale=0.7]{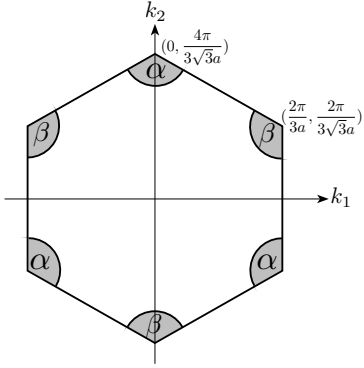}
\end{center}
\caption{\it Brillouin zone of the honeycomb lattice with corresponding hole
pockets.}
\end{figure}
The single-hole dispersion relation for the $t$-$J$ model on the honeycomb 
lattice is illustrated in figure 3.
\begin{figure}
\begin{center}
\includegraphics[scale=0.4,angle=-90]{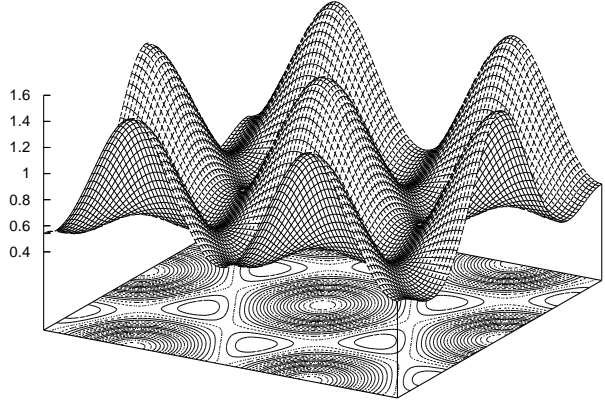}
\end{center}
\caption{\it Energy-momentum dispersion relation $E_h(k)/t$ for a single hole 
in the $t$-$J$ model on the honeycomb lattice for $J/t = 2$.}
\end{figure}

The effective field theory is defined in the space-time continuum and the
holes are described by Grassmann-valued fields $\psi^f_s(x)$ carrying a
"flavor" index $f = \alpha, \beta$ that characterizes the corresponding hole
pocket. The index $s = \pm$ denotes spin parallel ($+$) or antiparallel ($-$)
to the local staggered magnetization. Under the various symmetry operations
the hole fields transform as
\begin{eqnarray}
\label{trafoRules}
SU(2)_s: & & \psi^f_\pm(x)' = \exp(\pm i \alpha(x)) \psi^f_\pm(x),
\nonumber \\
U(1)_Q: & &^Q\psi^f_\pm(x) = \exp(i \omega) \psi^f_\pm(x),
\nonumber \\
D_i: & & ^{D_i}\psi^f_\pm(x) =
\exp(i k^f_i a_i) \psi^f_\pm(x),
\nonumber\\
O: & & ^O\psi^\alpha_\pm(x) = \mp \exp(\mp i \varphi(Ox)\pm i  
\frac{2 \pi}{3})\psi^\beta_\mp(Ox), \nonumber\\
& \quad & ^O\psi^\beta_\pm(x)= \mp \exp(\mp i \varphi(Ox)\mp i 
\frac{2 \pi}{3})\psi^\alpha_\mp(Ox), \nonumber \\
R: & & ^R\psi^\alpha_\pm(x) = \psi^\beta_\pm(Rx), \quad\;\;\;
^R\psi^\beta_\pm(x) = \psi^\alpha_\pm(Rx), \nonumber \\
T: & & ^T\psi^\alpha_\pm(x) = \exp(\mp i \varphi(Tx))
\psi^{\beta\dagger}_\pm(Tx),
\nonumber \\
& & ^T\psi^\beta_\pm(x) = \exp(\mp i \varphi(Tx))
\psi^{\alpha\dagger}_\pm(Tx),
\nonumber \\
& & ^T\psi^{\alpha\dagger}_\pm(x) = -\exp(\pm i \varphi(Tx)) 
\psi^\beta_\pm(Tx),\nonumber \\
&  & ^T\psi^{\beta\dagger}_\pm(x) = -\exp(\pm i \varphi(Tx)) 
\psi^\alpha_\pm(Tx).
\end{eqnarray}
Here $U(1)_Q$ is the fermion number symmetry of the holes. Interestingly, in 
the effective continuum theory the location of holes in lattice momentum space 
manifests itself as a "charge" $k^f_i$ under the displacement symmetry $D_i$.

Now that the relevant low-energy degrees of freedom have been identified and
the transformation rules of the corresponding fields have been worked out, the
construction of the effective action is uniquely determined. The low-energy
effective action of magnons and holes is constructed as a derivative
expansion. At low energies, terms with a small number of derivatives dominate
the dynamics. Since the holes are heavy nonrelativistic fermions, one 
time-derivative counts like two spatial derivatives. Here we limit ourselves
to terms with at most one temporal or two spatial derivatives. One then
constructs all terms consistent with the symmetries listed above. The
effective action can be written as
\begin{equation}
\label{effectiveaction}
S\left[\psi^{f\dagger}_\pm,\psi^f_\pm,v_\mu^\pm, v_\mu^3 \right] = \int d^2x \ dt
\ \sum_{n_\psi} {\cal L}_{n_\psi},
\end{equation} 
where $n_\psi$ denotes the number of fermion fields that the various terms 
contain. The leading terms in the pure magnon sector take the form
\begin{equation}\label{Lagrangian0} 
{\cal L}_0\, = 2\rho_s \left(v^+_i v^-_i + \frac{1}{c^2} v_t^+ v_t^-\right).
\end{equation}
The leading terms with two fermion fields (containing at most one temporal or
two spatial derivatives), describing the propagation of holes as well as their
couplings to magnons, are given by
\begin{eqnarray}
\label{Lagrangian2}
{\cal L}_2\,=\sum_{{f=\alpha,\beta};\,{\, s = +,-}}\Big[
& M \psi^{f\dagger}_s \psi^f_s + \psi^{f\dagger}_s D_t \psi^f_s
+ \frac{1}{2 M'} D_i \, \psi^{f\dagger}_s D_i \psi^f_s 
+\Lambda \psi^{f\dagger}_s (i s v^s_1 + \sigma_f v^s_2) \psi^f_{-s} \nonumber \\
& + iK\big[(D_1 + i s \sigma_f D_2) \psi^{f\dagger}_s
(v^s_1 + i s \sigma_f v^s_2)\psi^f_{-s} \nonumber \\
&\qquad-(v^s_1 + i s \sigma_f v^s_2)\psi^{f\dagger}_s (D_1 + i s  
\sigma_f D_2) \psi^f_{-s} \big] \nonumber \\
& +\sigma_f  L \psi^{f\dagger}_s \epsilon_{ij}f^3_{ij}\psi^f_s 
+ N_1 \psi^{f\dagger}_s v^s_i v^{-s}_i \psi^f_s \nonumber \\
& + i s \sigma_f N_2 \big( \psi^{f\dagger}_s v^s_1 v^{-s}_2  
\psi^f_s -
\psi^{f\dagger}_s v^s_2 v^{-s}_1 \psi^f_s \big) \Big].
\end{eqnarray}
Here $M$ is the rest mass and $M'$ is the kinetic mass of a hole, $\Lambda$
and $K$ are hole-one-magnon couplings, while $L$, $N_1$, and $N_2$ are
hole-two-magnon couplings. Note that all low-energy constants are real-valued.
The sign $\sigma_f$ is $+$ for $\alpha$ and $-$ for $\beta$. We have
introduced the field strength tensor of the composite Abelian "gauge" field
\begin{equation}
f^3_{ij}(x) = {\partial}_i v^3_j(x) - {\partial}_j v^3_i(x),
\end{equation}
and the covariant derivatives $D_t$ and $D_i$ acting on $\psi^{f}_{\pm}(x)$ as
\begin{eqnarray}
\label{kovardrhole}
D_t \psi^f_\pm(x) & = & \left[{\partial}_t \pm i v_t^3(x) - \mu \right]
\psi^f_\pm(x),
\nonumber \\
D_i \psi^f_\pm(x) &  = & \left[{\partial}_i \pm i v_i^3(x)\right] \psi^f_\pm(x).
\end{eqnarray}
The chemical potential $\mu$ enters the covariant time-derivative like an
imaginary constant vector potential for the fermion number symmetry $U(1)_Q$.
It is remarkable that the term proportional to $\Lambda$ with just a single
(uncontracted) spatial derivative satisfies all symmetries. Due to the small
number of derivatives it contains, this term dominates the low-energy dynamics
of a lightly hole-doped antiferromagnet on the honeycomb lattice.
Interestingly, for antiferromagnets on the square lattice, a corresponding
term, which was first identified by Shraiman and Siggia, is also present in
the hole-doped case \cite{Bru06}. On the other hand, a similar term is
forbidden by symmetry reasons in the electron-doped case \cite{Bru07b}. For
the honeycomb geometry we even identify a second hole-one-magnon coupling,
$K$, whose contribution, however, is sub-leading. Interestingly, hole- or
electron-doped antiferromagnets on the square lattice do not allow terms
containing the field-strength tensor $f_{ij}$ in ${\cal L}_2$.

Finally, the leading terms without derivatives and with four fermion fields
are given by
\begin{eqnarray}
\label{Lagrange4}
{\cal L}_4 & = & \sum_{s = +,-} \Big\{ \frac{G_1}{2} \, (\psi^{\alpha\dagger}_s
\psi^\alpha_s 
\psi^{\alpha\dagger}_{-s} \psi^\alpha_{-s} +
\psi^{\beta\dagger}_s \psi^\beta_s 
\psi^{\beta\dagger}_{-s} \psi^\beta_{-s}) \nonumber \\
& & + \ G_2 \, \psi^{\alpha\dagger}_s \psi^\alpha_s \psi^{\beta\dagger}_s
\psi^\beta_s 
+ G_3 \, \psi^{\alpha\dagger}_s \psi^\alpha_s 
\psi^{\beta\dagger}_{-s} \psi^\beta_{-s}\Big\}.
\end{eqnarray}
The low-energy four-fermion coupling constants $G_{1}$, $G_{2}$, and $G_{3}$
are also real-valued. Although potentially invariant under all symmetries,
terms with two identical hole fields vanish due to the Pauli principle.

It is important to note that in the above construction of the effective
Lagrangian for the $t$-$J$ model which contains holes only, a crucial step was
to identify the degrees of freedom that correspond to the holes. In order to
remove the electron degrees of freedom one has to explicitly break the
particle-hole $SU(2)_Q$ symmetry, leaving the ordinary fermion number symmetry
$U(1)_Q$ intact. This task can be achieved by constructing all possible
fermionic mass terms that are invariant under the various symmetries. Picking
the eigenvectors which correspond to the lowest eigenvalues of the mass
matrices then allows one to separate electrons from holes. This procedure is
described in detail in Ref.~\cite{KBWHJW12}

We would like to point out that the leading order terms in the effective
Lagrangian for magnons and holes constructed above exhibit two accidental
global symmetries. First, we notice that for $c\rightarrow\infty$ and without
the term proportional to $iK$ in $\mathcal L_{2}$, Eqs.(\ref{Lagrangian0}) and
(\ref{Lagrangian2}), have an accidental Galilean boost symmetry. Although the
Galilean boost symmetry is explicitly broken at higher orders of the
derivative expansion, this symmetry has physical implications: The leading
one-magnon exchange between two holes, to be discussed in the next section,
can be investigated in their rest frame without loss of generality.

In addition, we notice an accidental global rotation symmetry $O(\gamma)$.
Except for the term proportional to $iK$, $\mathcal L_{2}$ of
Eq.(\ref{Lagrangian2}) is invariant under a continuous spatial rotation by an
angle $\gamma$. This symmetry is not present in the $\Lambda$-term of the
square lattice. The $O(\gamma)$ invariance has some interesting implications
for the spiral phases in a lightly doped antiferromagnet on the honeycomb
lattice, as we will see in the next section. 

\section{Applications}

In this section we will consider two nontrivial applications of the effective
Lagrangian constructed above: The emergence of spiral phases in the staggered
magnetization vector order parameter and the formation of two-hole bound
states.

\subsection{Spiral phases}

Let us first investigate whether so-called spiral phases in the staggered
magnetization can occur upon hole--doping. This problem has been studied
before using microscopic theories \cite{Shr88,Shr92,Sus04,Kot04}. In
Ref.~\cite{Bru07a} the problem was studied systematically in the case of a
square lattice using effective Lagrangians. Here we consider the honeycomb
lattice -- details of the calculation can be found in Ref.~\cite{JKHW09}. 

In our study we assume that the holes are doped into the system homogeneously
and thus limit ourselves to configurations of the staggered magnetization that
are either homogeneous themselves or generate a constant background field
$v_i(x)$ for the charge carriers. As was shown in \cite{Bru07a}, the most
general configuration of this kind represents a spiral in the staggered
magnetization. We also assume that four-fermion contact interactions are weak
and take them into account perturbatively -- again, for details see
Refs.\cite{Bru07b,Bru07a}.

When one populates the various hole pockets, one must distinguish a total of
four different cases. First we have $\alpha$-pockets and $\beta$-pockets.
Moreover, the energy  eigenstates of a hole pocket $f$ can acquire two values,
$E^f_{-}$ and  $E^f_{+}$, where the former corresponds to the lower energy. One
thus starts considering the case of populating all four hole pockets, i.e.\
with both flavors $f = \alpha, \beta$  and with both energy indices $\pm$. One
then proceeds with populating only three, only two and, finally, only one hole
pocket. This gives rise to various phases of the staggered magnetization,
which can be either homogeneous or a spiral phase as depicted in figure 4.

\begin{figure}
\begin{center}
\includegraphics[scale=0.6]{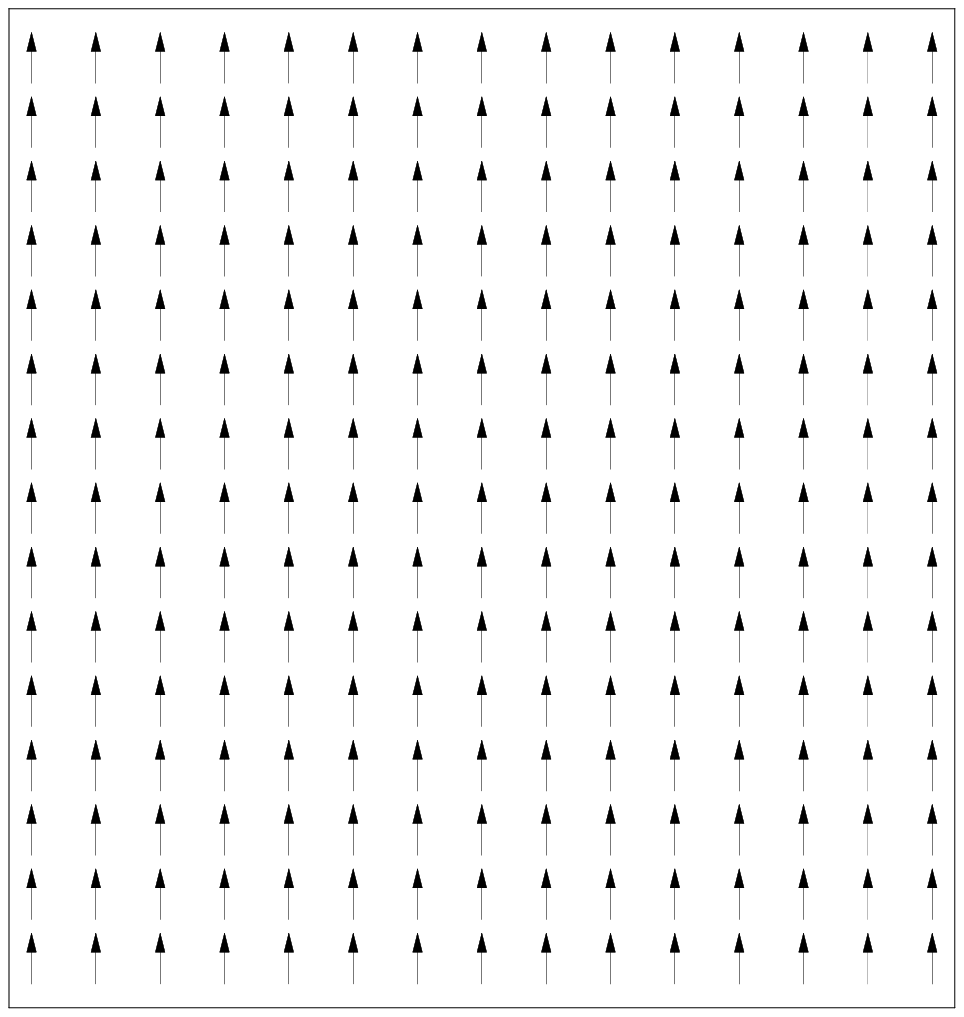}
\includegraphics[scale=0.65]{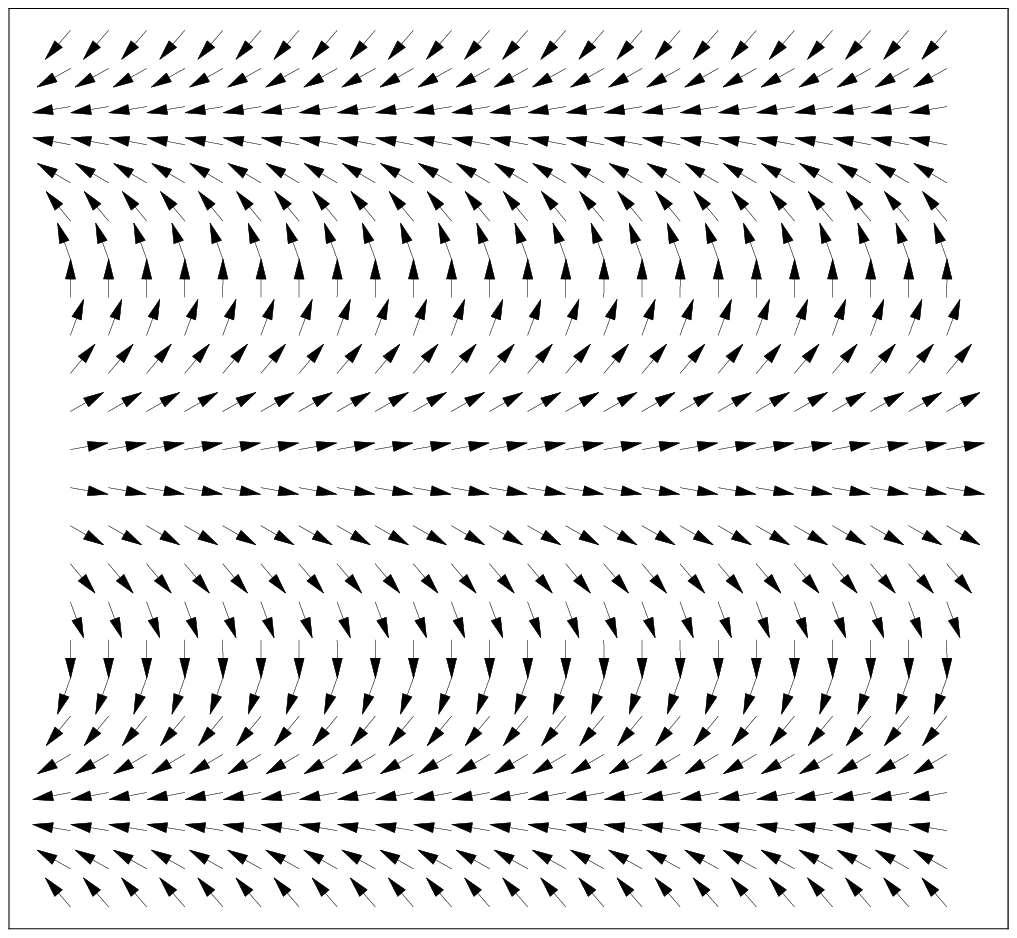}
\caption{Left panel: Homogeneous phase with constant staggered magnetization.
Right panel: Spiral phase with helical structure in the staggered
magnetization.}
\end{center}
\end{figure}

More precisely, the energy densities of the various phases take the form
\begin{equation}
\epsilon_i = \epsilon_0 + M n + \frac{1}{2} \kappa_i n^2.
\end{equation}
Here $\epsilon_0$ is the energy density of the system at half-filling and $n$
is the total density of holes. The index $i$ refers to the number of hole
pockets that are populated in the corresponding phase. The compressibilities
$\kappa_i$ are given by
\begin{eqnarray}
\kappa_1&=&\frac{2 \pi}{M'} - \frac{\Lambda^2}{4 \rho_s}, \nonumber \\
\kappa_2&=&\frac{\pi}{M'} - \frac{\Lambda^2}{4 \rho_s}
+ \frac{1}{4} (G_2 + G_3), \nonumber \\
\kappa_3&=&\frac{2 \pi}{3 M'} \left(1 - \frac{1}{8}\frac{M' \Lambda^2}
{3 \pi \rho_s - M' \Lambda^2}\right) \nonumber \\
&+&\frac{4 \pi \rho_s - M' \Lambda^2}{(3 \pi \rho_s - M' \Lambda^2)^2}
\frac{1}{16} \Big[8(G_1 + G_2 + G_3)\pi\rho_s \Big. \nonumber \\
&-&\Big. (4 G_1+ 3 G_2 + 3 G_3) M' \Lambda^2\Big], \nonumber \\
\kappa_4&=&\frac{\pi}{2 M'} + \frac{1}{4} (G_1 + G_2 + G_3),
\end{eqnarray}
and shown in figure 5 as functions of $M'\Lambda^2/2\pi\rho_s$. 
\begin{figure}
\begin{center}
\includegraphics[scale=0.5]{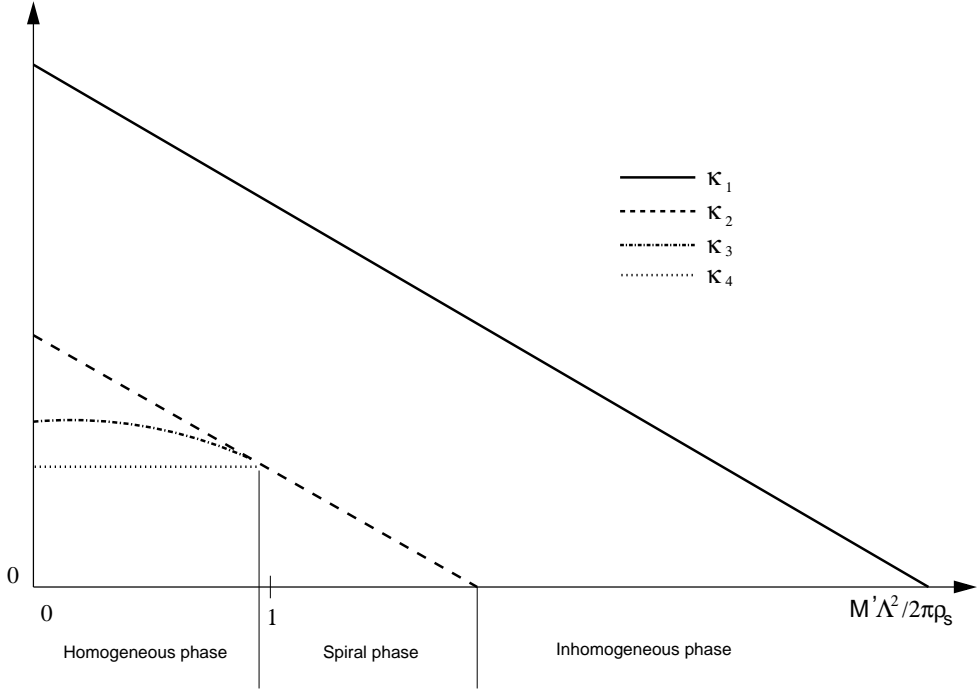}
\end{center}
\caption{\it The compressibilities $\kappa_i$ as functions of 
$M' \Lambda^2/2 \pi \rho_s$ determine the stability ranges of the various 
phases. A homogeneous phase, a spiral, or an inhomogeneous phase are
energetically favorable, for large, intermediate, and small values of
$\rho_s$, respectively.}
\end{figure}
For large values of $\rho_s$, spiral phases cost a large amount of magnetic 
energy and the homogeneous phase is more stable. To be more precise, in this 
regime one has $\kappa_4 < \kappa_3 < \kappa_2 < \kappa_1$. Notice that
$\kappa_1$ is always larger than $\kappa_2$ for any value of $\rho_s$. As
$\rho_s$ decreases and reaches the value
\begin{equation}
\label{critical}
\rho_s = \frac{M'\Lambda^2}{2\pi} + \frac{(M')^{2}\Lambda^2G_1}{4\pi^2},
\end{equation}
at leading order in the 4-fermi couplings one finds
$\kappa_2 = \kappa_3 = \kappa_4$. For smaller values of $\rho_s$, the
two-pocket spiral is energetically favored until $\kappa_2$ becomes negative
and the system becomes unstable against the formation of spatial
inhomogeneities of a yet undetermined type.

Interestingly, unlike in the square lattice case, due to the accidental
continuous $O(\gamma)$ spatial rotation symmetry, at leading order a spiral
does not have an a priori preferred spatial direction. It is instructive to
compare the results presented  here with the results obtained in the square
lattice case \cite{Bru07a}. Qualitatively the stability ranges of various
phases are the same for both lattice geometries except that the one-pocket
spiral is never energetically favored on the honeycomb lattice while it is
favorable in a small parameter regime on the square lattice.  

\subsection{Two-hole bound states}

In the effective theory framework, at low energies, holes interact with each
other via magnon exchange. Since the long-range dynamics is dominated by
one-magnon exchange, we will calculate the one-magnon exchange potentials
between two holes of the same flavor $\alpha$ and $\beta$ and of different
flavor, and then address the question regarding the formation of two-hole
bound states. The same problem was also considered in
Refs.~\cite{Sus93,Kuc93,Fla94} using microscopic theories. 

We expand in the magnon fluctuations $m_1(x)$ and $m_2(x)$ around the ordered
staggered magnetization
\begin{equation}
\vec e(x) = \left( \frac{m_1(x)}{\sqrt{\rho_s}},\,
\frac{m_2(x)}{\sqrt{\rho_s}},1 \right) + {\cal O}\left(m^2\right).
\end{equation}
For the composite magnon fields this implies
\begin{eqnarray}
v_\mu^\pm(x) & = & \frac{1}{2 \sqrt{\rho_s}} {\partial}_\mu
\big[ m_2(x) \pm i m_1(x) \big] + {\cal O}\left(m^3\right), \nonumber \\
v_\mu^3(x) & = & \frac{1}{4 \rho_s}\big[m_1(x) {\partial}_\mu m_2(x) -
m_2(x) {\partial}_\mu m_1(x)\big] + {\cal O}\left(m^4\right).
\end{eqnarray}
Vertices with $v_\mu^3(x)$ involve at least two magnons, such that one-magnon
exchange results from vertices with $v_\mu^\pm(x)$ only. As a consequence, two
holes can exchange a single magnon only if they have anti-parallel spins ($+$
and $-$), which are both flipped in the magnon-exchange process. We denote the
momenta of the incoming and outgoing holes by $\vec p_\pm$ and $\vec p_\pm\!'$,
respectively. The momentum carried by the exchanged magnon is denoted by
$\vec q$. The incoming and outgoing holes are asymptotic free particles with
momentum $\vec p=(p_1,p_2)$ and energy $E(\vec p)= M +  p_i^2/2M'$. One-magnon
exchange between two holes is associated with the Feynman diagram in
figure 6.
\begin{figure}
\begin{center}
\vspace{-0.4cm}
\includegraphics[width=7cm]{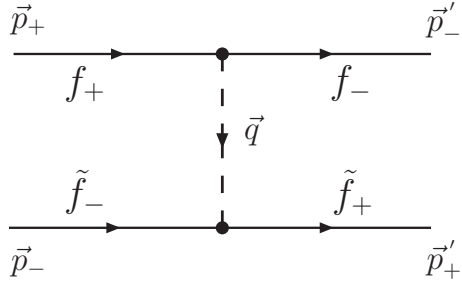}
\end{center}
\caption{\it Tree-level Feynman diagram for one-magnon exchange between two
holes.}
\end{figure}
Evaluating these Feynman diagrams, one finds the resulting potentials for the
various combinations of flavors $f, \tilde f \in \{\alpha, \beta\}$ and
couplings $F,\tilde F \in\{\Lambda,K\}$.

The leading contribution to the low-energy physics, as we noted earlier, comes
from the $\Lambda$ vertex. From here on, we therefore concentrate on the
potential with two $\Lambda$ vertices only. In coordinate space the
$\Lambda\Lambda$-potentials are given by
\begin{equation}\label{general_potentials}
\langle \vec r_+ \ \!\!\!\! ' \vec r_- \ \!\!\!\! '|V^{f\tilde f}_{\Lambda\Lambda}
|\vec r_+ \vec r_-\rangle = V^{f\tilde f}_{\Lambda\Lambda}(\vec r\, ) \
\delta(\vec r_+ - \vec r_- \ \!\!\!\! ') \
\delta(\vec r_- - \vec r_+ \ \!\!\!\! '),
\end{equation}
with
\begin{equation}
V^{ff}_{\Lambda\Lambda}(\vec r\,) = -\frac{\Lambda^2}{2\rho_s}\delta^{(2)}(\vec r
\,),\qquad
V^{ff'}_{\Lambda\Lambda}(\vec r\,) = \frac{\Lambda^2}{2\pi \rho_s\vec r^{\,2}}
\exp (2i\sigma_f \varphi).
\end{equation}
Here $\vec r = \vec r_+ - \vec r_-$ denotes the distance vector between the
two holes and $\varphi$ is the angle between $\vec r$ and the $x_1$-axis. The
\mbox{$\delta$-f}unctions in Eq.(\ref{general_potentials}) ensure that the
holes do not change their position during the magnon exchange. Note that the
one-magnon exchange potentials are instantaneous although magnons travel with
the finite speed $c$. Retardation effects occur only at higher orders. 

Interestingly, in the $\Lambda\Lambda$ channel, one-magnon exchange over long
distances between two holes can only happen for holes of opposite flavor. For
two holes of the same flavor, one-magnon exchange acts as a contact
interaction. Here we concentrate on the long-range physics of weakly bound
states of holes and therefore will only consider the binding of holes of
different flavor. 

Let us therefore investigate the Schr\"odinger equation for the relative
motion of two holes with flavors $\alpha$ and $\beta$. In the following, we
will take care of short distance interactions by imposing a hard-core boundary
condition on the pair's wave function. Due to the accidental Galilean boost
invariance, without loss of generality, we can consider the hole pair in its
rest frame. The total kinetic energy of the two holes is given by
\begin{equation}
T =  \sum_{f=\alpha,\beta} T^{f} = \sum_{f = \alpha,\beta} \frac{p_i^2}{2 M'}
= \frac{p_i^2}{M'}. 
\end{equation}
We now introduce the two probability amplitudes \mbox{$\Psi_1(\vec r \,)$} and
\mbox{$\Psi_2(\vec r \,)$} which represent the two flavor-spin combinations
\mbox{$\alpha_+\beta_-$} and \mbox{$\alpha_-\beta_+$}, respectively, with the
distance vector \mbox{$\vec r$} to point from the \mbox{$\beta$} to the
\mbox{$\alpha$} hole. Because the holes undergo a spin flip during the magnon
exchange, the two probability amplitudes are coupled through the magnon
exchange potentials and the Schr\"odinger equation describing the relative
motion of the hole pair is a two-component equation. Using the explicit form
of the potentials, it takes the form
\begin{equation}
\label{schroedinger}
\left(\begin{array}{cc} - \frac{1}{M'}\Delta & \gamma\frac{1}{\vec r^{\,2}}
\exp(-2i\varphi)
\\[0.2ex]
\gamma\frac{1}{\vec r^{\,2}}\exp(2i\varphi) &  - \frac{1}{M'} \Delta 
\end{array} \right)
\left(\begin{array}{c} \Psi_1(\vec r \, ) \\ 
\Psi_2(\vec r \, ) \end{array}\right) = E 
\left(\begin{array}{c} \Psi_1(\vec r \, ) \\ 
\Psi_2(\vec r \, ) \end{array}\right),
\end{equation}
with
\begin{equation}\label{gamma}
\gamma = \frac{\Lambda^2}{2\pi \rho_s}.
\end{equation}
As it turns out, magnon-mediated forces can lead to bound states only if the
low-energy constant $\Lambda$ is larger than a critical value given by
\begin{equation}
\Lambda_c = \sqrt{\frac{2\pi \rho_s}{M'}}.
\end{equation}
Interestingly, the same critical value also arose in the investigation of
spiral phases in a lightly doped antiferromagnet on the honeycomb lattice
in Eq.(\ref{critical}). There it marked the point where spiral phases become
energetically favorable compared to the homogeneous phase.

With the separation ansatz
\begin{equation}
\label{ansatz}
\Psi_1(r,\varphi) = R_1(r) \exp (i m_1 \varphi), \qquad \Psi_2(r,\varphi) =
R_2(r) \exp (i m_2 \varphi),
\end{equation}
the relevant equation which is associated with an attractive potential and
thus potentially leads to the formation of bound states amounts to
\begin{equation}
\label{radialeq}
\left[-\left(\frac{d^2}{d r^2}+\frac{1}{r}\frac{d}{d r}\right) +(1 -\gamma M')
\frac{1}{r^{2}}\right]R(r) = - M' |E| R(r),
\end{equation}
with $E=-|E|$ and $R(r)=R_1(r)-R_2(r)$. It has to be pointed out that this
equation refers to the case $m_1=-1$ and $m_2=1$ where the system
(\ref{schroedinger}) can be decoupled (see \cite{KBWHJW12}).

The same equation also occurred in the square lattice case \cite{Bru06,Bru06a}
and can be solved along the same lines. As it stands, the equation is
ill-defined because the $1/r^2$ potential is too singular at the origin.
However, we have not yet included the contact interaction proportional to the
4-fermion coupling $G_3$. Here, in order to keep the calculation analytically
feasible, we model the short-range repulsion by a hard core radius $r_0$,
i.e.\ we require \mbox{$R(r_0)=0$} for \mbox{$r \leq r_0$}.
Eq.(\ref{radialeq}) is solved by a modified Bessel function
\begin{equation}
R(r) = A K_\nu \big( \sqrt{M' |E|} r \big), \qquad \nu = i \sqrt{\gamma M' -1},
\end{equation}
with $A$ being a normalization constant. Demanding that the wave function
vanishes at the hard core radius gives a quantization condition for the bound
state energy. The quantum number $n$ then labels the $n$-th excited state. For
large $n$, the binding energy is given by
\begin{equation}
E_n \sim - \frac{1}{M' r_0^2} \exp\left(\frac{- 2 \pi n}{\sqrt{\gamma M' -1}}
\right).
\end{equation}
The binding is exponentially small in $n$ and there are infinitely many bound
states. While the highly excited states have exponentially small energy, for
sufficiently small $r_0$ the ground state could have a small size and be
strongly bound. However, for short-distance physics the effective theory
should not be trusted quantitatively. Still, as long as the binding energy is
small compared to the relevant high-energy scales, our result is valid and
receives only small corrections from higher-order effects such as two-magnon
exchange.

We now turn to the discussion of the angular part of the wave equation. The
ansatz (\ref{ansatz}) leads to the following solution for the ground state
wave function
\begin{equation}
\label{groundstatewavefunction}
\Psi(r, \varphi) = 
\left(\begin{array}{c} \Psi_1(\vec r \, ) \\ 
\Psi_2(\vec r \, ) \end{array}\right)
= R(r) \, 
\left(\begin{array}{c} \exp(- i \varphi) \\ 
- \exp(i \varphi) \end{array}\right) .
\end{equation}
Under the 60 degrees rotation $O$, using the transformation rules of
Eq.(\ref{trafoRules}), one obtains
\begin{equation}
^O\Psi(r, \varphi) = - \, \Psi(r, \varphi).
\end{equation}
Interestingly, the wave function for the ground state of two holes of flavors
$\alpha$ and $\beta$ thus exhibits $f$-wave symmetry.\footnote{Strictly 
speaking, the continuum classification scheme of angular momentum eigenstates 
does not apply here, since we are not dealing with a continuous rotation
symmetry.} The corresponding probability distribution depicted in figure 7, on
the other hand, seems to show $s$-wave symmetry. However, the relevant phase
information is not visible in this picture, because only the probability
density is shown.
\begin{figure}
\begin{center}
\vspace{-0.4cm}
\includegraphics[width=6cm]{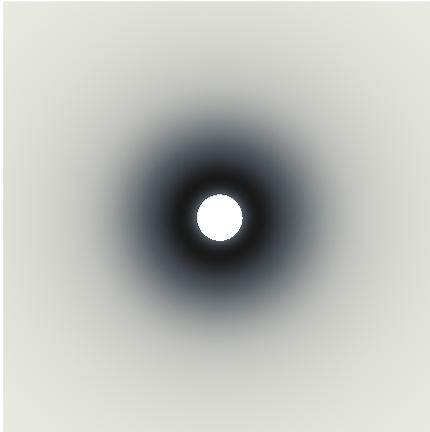}
\end{center}
\caption{\it Probability distribution for the ground state of two holes of
flavors $\alpha$ and $\beta$.}
\end{figure}
Interestingly, for two-hole bound states on the square lattice, the wave 
function for the the ground state of two holes of flavors $\alpha$ and $\beta$ 
shows $p$-wave symmetry, while the corresponding probability distribution
(which again does not contain the relevant phase information) resembles
$d_{x^2-y^2}$ symmetry \cite{Bru06}. Remarkably, the ground state wave function 
(\ref{groundstatewavefunction}) of a bound hole pair on the honeycomb lattice 
remains invariant under the reflection symmetry $R$, the shift symmetries 
$D_i$, as well as under the accidental continuous rotation symmetry
$O(\gamma)$.

We find it quite remarkable that the $f$-wave character of the two-hole bound 
state on the honeycomb lattice is an immediate consequence of our systematic 
effective field theory analysis. The question regarding the true symmetry of
the pairing state realized in the dehydrated version of Na$_2$CoO$_2 \times y$
H$_2$O, still seems to be controversial \cite{Iva09}. Still, a careful
analysis of the available experimental data for this compound suggests that
the pairing symmetry indeed is $f$-wave \cite{Maz05}.

\section{Conclusions}

The effective theory for the insulating antiferromagnetic precursors of
high-temperature superconductors is the condensed matter analog of baryon
chiral perturbation theory: Magnons play the role of the pseudoscalar mesons,
while the holes and electrons are the analog of the baryon octet. We have
analyzed the symmetries of the underlying Hubbard and $t-J$-type models in
detail and reviewed the systematic construction of the effective low-energy
field theory for weakly hole-doped antiferromagnets on the honeycomb lattice.
The procedure is fully systematic order by order in a derivative expansion of
the magnon and hole fields. As two nontrivial applications we have considered
the existence of spiral phases in the staggered magnetization order parameter
and the formation of two-hole bound states mediated by magnon exchange.

Regarding spiral phases, we have limited ourselves to constant composite
vector fields $v_i(x)'$ which implies that the fermions experience a constant
background field. Unlike in the square lattice case, due to the accidental
continuous $O(\gamma)$ spatial rotation symmetry, at leading order a spiral
does not have an a priori preferred spatial direction. However, since the
$O(\gamma)$ symmetry is broken explicitly by the higher-order terms, once such
terms are included, one expects the spiral to align with a  lattice direction.
We also investigated the stability of spiral phases in the presence of
4-fermion couplings. If these couplings can be treated perturbatively, for
sufficiently large values of $\rho_s$, the homogeneous phase is energetically
favored. With decreasing $\rho_s$, a two-pocket spiral becomes energetically
more favorable. In contrast to the square lattice case, the one-pocket spiral
is never favored. For small values of $\rho_s$ the two-pocket spiral becomes
unstable against the formation of inhomogeneities of a yet undetermined type.

Regarding the formation of two-hole bound states mediated by magnon-exchange,
we have studied the effect of the magnon-hole vertex. Again in contrast to the
square lattice case, it turned out that the magnon-hole coupling constant
$\Lambda$ must exceed a critical value in order to obtain two-hole bound
states. Our analysis implies that the wave function for the ground state of
two holes of flavors $\alpha$ and $\beta$ exhibits $f$-wave symmetry (while
the corresponding probability distribution seems to suggest $s$-wave
symmetry). This is quite different from the square lattice case, where the
wave function for the ground state of two holes of  flavors $\alpha$ and
$\beta$ exhibits $p$-wave symmetry (while the corresponding probability
distribution resembles $d_{x^2-y^2}$ symmetry).

It is quite remarkable that all these results unambiguously follow from the
very few basic assumptions of the systematic low-energy effective theory, such
as symmetry, locality, and unitarity. The effective theory provides a
theoretical framework in which the low-energy dynamics of lightly hole-doped
antiferromagnets can be investigated in a systematic manner. In particular,
after the low-energy parameters have been adjusted appropriately, the
resulting low-energy physics is completely equivalent to the one of the
Hubbard or $t$-$J$ model.

\ack C.P.H. would like to thank the members of the Institute for Theoretical
Physics at Bern University for their warm hospitality during a visit at which
this project was completed. Support by CONACYT grant No. 50744-F is gratefully
acknowledged.

\end{document}